\documentclass[aps,prl,reprint,superscriptaddress]{revtex4-1}
\usepackage{amsmath,braket}
\usepackage{graphicx}

\usepackage{color}
\usepackage{hyperref}

\usepackage[separate-uncertainty=true,multi-part-units=single]{siunitx}

\definecolor{darkgreen}{rgb}{0.0,0.4,0.0}
\DeclareSIUnit{\mu}{\micro\meter}
\DeclareSIUnit{\unit}{\relax}

\begin{document}


\title{Demonstrating the decoupling regime of the electron-phonon interaction in a quantum dot using chirped optical excitation}



\author{Timo Kaldewey}
\affiliation{Department of Physics, University of Basel, Klingelbergstrasse 82, CH-4056 Basel, Switzerland}

\author{Sebastian L\"{u}ker}
\affiliation{Institut f\"{u}r Festk\"{o}rpertheorie, Universit\"{a}t M\"{u}nster, Wilhelm-Klemm-Strasse 10, D-48149 M\"{u}nster, Germany}

\author{Andreas V.\ Kuhlmann}
\affiliation{Department of Physics, University of Basel, Klingelbergstrasse 82, CH-4056 Basel, Switzerland}
\affiliation{IBM Research-Zurich, S\"{a}umerstrasse 4, CH-8803 R\"{u}schlikon, Switzerland}

\author{Sascha R.\ Valentin}
\affiliation{Lehrstuhl f\"{u}r Angewandte Festk\"{o}rperphysik, Ruhr-Universit\"{a}t Bochum, D-44780 Bochum, Germany}

\author{Jean-Michel Chauveau}
\affiliation{ Universit\'{e} C\^{o}te d'Azur, CNRS, CRHEA, France}

\author{Arne Ludwig}
\affiliation{Lehrstuhl f\"{u}r Angewandte Festk\"{o}rperphysik, Ruhr-Universit\"{a}t Bochum, D-44780 Bochum, Germany}

\author{Andreas D.\ Wieck}
\affiliation{Lehrstuhl f\"{u}r Angewandte Festk\"{o}rperphysik, Ruhr-Universit\"{a}t Bochum, D-44780 Bochum, Germany}

\author{Doris E. Reiter}
\affiliation{Institut f\"{u}r Festk\"{o}rpertheorie, Universit\"{a}t M\"{u}nster, Wilhelm-Klemm-Strasse 10, D-48149 M\"{u}nster, Germany}

\author{Tilmann Kuhn}
\affiliation{Institut f\"{u}r Festk\"{o}rpertheorie, Universit\"{a}t M\"{u}nster, Wilhelm-Klemm-Strasse 10, D-48149 M\"{u}nster, Germany}

\author{Richard J.\ Warburton}
\affiliation{Department of Physics, University of Basel, Klingelbergstrasse 82, CH-4056 Basel, Switzerland}



\date{\today}

\begin{abstract}
Excitation of a semiconductor quantum dot with a chirped laser pulse allows excitons to be created by rapid adiabatic passage. In quantum dots this process can be greatly hindered by the coupling to phonons. Here we add a high chirp rate to ultra-short laser pulses and use these pulses to excite a single quantum dot. We demonstrate that we enter a regime where the exciton-phonon coupling is effective for small pulse areas, while for higher pulse areas a decoupling of the exciton from the phonons occurs. We thus discover a reappearance of rapid adiabatic passage, in analogy to the predicted reappearance of Rabi rotations at high pulse areas. The measured results are in good agreement with theoretical calculations.
\end{abstract}


\maketitle 

In semiconductors, a driven electron is damped by the interaction with phonons. In the context of quantum control, phonons therefore lead to dephasing. The electron-phonon interaction is therefore important in the development of quantum technology with semiconductors. It is a rich and subtle subject.

One possible way to suppress electron-phonon damping is to drive the electronic system so quickly that the relatively large inertia of the phonons presents them from reacting to the driven electron. In the context of Rabi oscillations, the driven oscillations of a two-level system, a ``reappearance" has been predicted \cite{vagov2007non}. As the drive is increased, the Rabi oscillations are initially damped more and more by the phonons but then the damping decreases and is eventually suppressed. The reappearance regime represents phonon-free quantum control. It has however never been observed experimentally. Here, we demonstrate the experimental realization of the “reappearance regime”. Validation comes from a full microscopic theory.

Our quantum system is a single self-assembled quantum dot (QD), an emitter of highly coherent single photons and polarization-entangled photon pairs \cite{Kaldewey2017biexc}. Quantum control of the exciton, an electron-hole pair, proceeds on picosecond time-scales well before spontaneous emission takes place (time-scale \SI{\sim1}{\nano\second}). Phonons lead to a deterioration of the exciton preparation fidelity for schemes using resonant excitation \cite{Ramsay2010_oct,Ramsay2010,Ramsay2010review,krugel2005the,vagov2007non,Reiter2014}. In fact, the interaction with the phonons is sufficiently strong that an exciton state can be prepared by relying on it (phonon-mediated relaxation following excitation with a detuned pulse) \cite{Reiter2012,Glassl2013Phonon,Quilter2015pho,Bounouar2015,Ardelt2014}. In a Rabi experiment, phonons lead to a clear damping \cite{Ramsay2010_oct,Ramsay2010}. The specific dephasing mechanism was identified as a coupling to longitudinal acoustic (LA) phonons. For higher pulse areas, theory predicts that the electronic oscillations become so fast such that the phonons decouple and the Rabi oscillations recover, the reappearance phenomenon. The existence of a pulse area for which the coupling to the phonons is maximal is a consequence of the non-monotonic electron-phonon coupling \cite{vagov2007non,wigger2014ene}.

For the pulses used so far experimentally (pulses of 1-10~ps duration), the reappearance regime for Rabi oscillations can only be entered at extremely high pulses areas ($>20 \pi$) and has therefore remained out of reach. We switch to an alternative technique here, rapid adiabatic passage (RAP) \cite{Simon2011,Wu2011,Luker2012,Gawarecki2012,debnath2012chi,Glassl2013,Mathew2014,Wei2014} and use the full bandwidth of \SI{100}{\femto\second} pulses. There are two key advantages. First, for such short laser pules, the reappearance regime moves to lower pulse areas which are easier to access experimentally. Secondly, low energy phonons do not contribute to the damping in the RAP process. Technically, this arises because there is always a finite splitting between the dressed states. Conversely, in a Rabi oscillation with Gaussian pulses, the splitting between the dressed states increases monotonically from zero such that the full range of phonons is involved in the damping. This second feature also makes the reappearance regime easier to access.

Rapid adiabatic passage (RAP) has been demonstrated on single QDs \cite{Simon2011,Wu2011,Wei2014} and it has been shown that phonons hinder exciton preparation depending on the sign of the chirp \cite{Luker2012,Mathew2014}. It has been predicted that the reappearance regime translates to a non-monotonic RAP behavior: at sufficiently high pulse areas RAP improves \cite{Reiter2012}. We demonstrate exactly this improvement here by comparing RAP experiments with a full microscopic theory of exciton-phonon dephasing. We therefore present compelling evidence that our system enters the reappearance regime where the electron is effectively decoupled from the phonons. We show that RAP is excellent for phonon-free state preparation.

\begin{figure}[t]{}
	\centering
	\includegraphics[width=1\columnwidth]{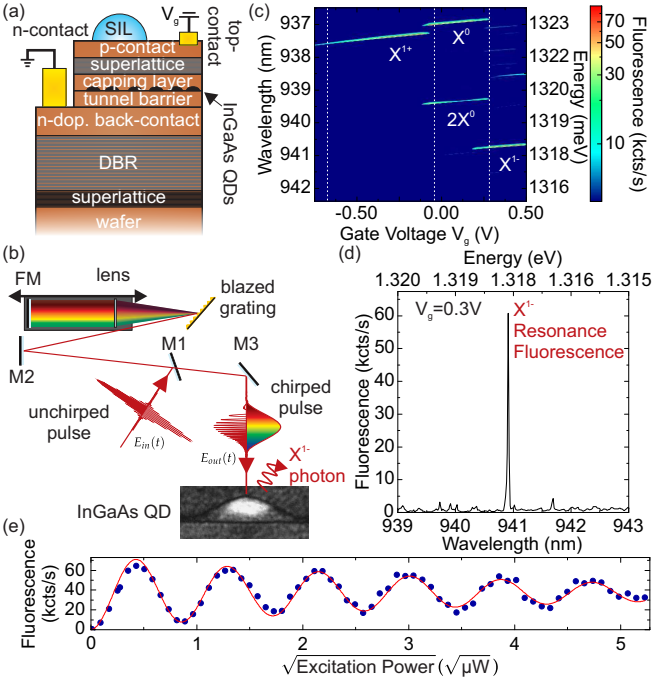}
	\caption{(a) Scheme of the n-i-p structure with an embedded layer of quantum dots (QDs).
	(b) Scheme of the folded $4f$ pulse-shaper to control the chirp introduced into an ultra-short, transform-limited laser pulse. The unchirped pulse is directed with mirrors (M) onto a grating, focused onto a folding mirror (FM) and back-reflected with a slight angle overshooting M1. The chirped pulse is then sent to the microscope (not shown) and excites the QD (TEM image, 28 x 8 \si{\nano\meter\squared}) which emits a resonance fluorescence photon.
	(c) Response of the QD to broadband excitation as a function of $V_g$. A clear Coulomb blockade is observed. The excitonic transitions are identified. 	The excitation pulses (linear polarization, \SI{938.22}{\nano\meter} center-wavelength) were positively chirped.
	(d) Detected resonance fluorescence signal after broadband excitation as a function of the detection wavelength. The peak arises from emission from the $\ket{\mathrm{X}^{1-}}\rightarrow\ket{\mathrm{e}^{1-}}$ transition. The gate voltage was $V_g=\SI{0.3}{\volt}$ at which a single electron resides in the QD.
	(e) Rabi rotations driven on the $\ket{\mathrm{X}^{1-}}\leftrightarrow\ket{\mathrm{e}^{1-}}$ transition with a \SI{2}{\pico\second} long, transform-limited pulse. Blue points show the detected resonance fluorescence signal, red curve is a damped sine fit.
	}
	\label{fig:scheme}
\end{figure}

We study a self-assembled InGaAs QD embedded in an n-i-p structure as displayed in Fig.\ \ref{fig:scheme}(a) and described in the supplementary information (SI) \cite{SI}. By applying a gate voltage of \SI{0.3}{\volt}, the QD is occupied by a single electron such that the QD mimics a two-level system. The ground state is the single electron state, $\ket{\mathrm{e}^{1-}}$, and the excited state the negatively-charged trion, $\ket{\mathrm{X}^{1-}}$. The QD is excited using chirped laser pulses. A mode-locked laser produces transform-limited pulses which are then manipulated in a folded $4f$ pulse-shaper \cite{Martinez1987}. The pulses have an intensity full-width-at-half-maximum (FWHM) of $\Delta t_\mathrm{FWHM}=\SI{130}{\femto\second}$ with close-to-Gaussian pulse form with $\Omega_0(t)=\frac{\Theta}{\tau_0	\sqrt{2\pi}}\exp\left(-\frac{t^2}{2\tau_0^2}\right)\exp\left(-i\omega_L t\right)$, where $\Theta$ is the pulse area, $\tau_0=\Delta t_\mathrm{FWHM}/(2\sqrt{\ln{2}})$ is the pulse width and $\omega_L$ the center frequency. The Rabi frequency $\Omega$ is related to the electric field $\mathbf{E}$ of the laser pulse by the dipole matrix element $\mathbf{M}$ via $\hbar\Omega(t)=2\mathbf{M}\cdot \mathbf{E}(t)$.
The pulse-shaper adds a frequency dependent phase resulting in a chirp coefficient $\alpha$. The chirp stretches the pulse in time to $\tau=\sqrt{\frac{\alpha^2}{\tau_0^2}+\tau_0^2}$. The instantaneous frequency of the laser pulse changes in time with the frequency chirp rate $a=\mathrm{d}\omega/\mathrm{d}t=\frac{\alpha}{\alpha^2+\tau_0^4}$ \cite{Malinovsky2001}, such that after the pulse shaper the pulse reads
\begin{equation}
\Omega(t)=\frac{\Theta}{\sqrt{2\pi\tau_0\tau}}\exp\left(-\frac{t^2}{2\tau^2}\right)
\exp\left(-i(\omega_L+\frac{1}{2}a t) t\right).
\end{equation}
The central wavelength ($2\pi c/\omega_L$) of the pulses is detuned by \SI{2.58}{\nano\meter} from the trion transition corresponding to an energy of \SI{3.63}{\milli\electronvolt}. A chirp coefficient $\alpha$ of \SI{0.31}{\pico\second\squared} (\SI{0.66}{\pico\second\squared}) stretches the pulse length to $7$~ps ($15$~ps). A sketch of the experiment is shown in Fig.~\ref{fig:scheme}(b): the QD at temperature $T=4.2$~K is excited with a chirped pulse and the resonance fluorescence is detected. The full set-up is described in the SI \cite{SI}. A cross-polarized dark-field technique suppresses the reflected laser light from the detection channel \cite{Vamivakas2009,Ylmaz2010,Kuhlmann2013_RSI,Kuhlmann2013_NatPhys}; further rejection of the laser light is carried out with a grating spectrometer (the laser pulse is broadband, the resonance fluorescence narrowband). The device works well even when excited with the broadband laser pulses. The Coulomb blockade is robust: Fig.~\ref{fig:scheme}(c) shows the resonance fluorescence response to a broadband excitation as a function of the gate voltage (see also SI \cite{SI}). Within the $\mathrm{X}^{1-}$ plateau, resonance fluorescence emerges just from the $\mathrm{X}^{1-}$ validating the two-level assertion. For an excitation power of \SI{0.9}{\micro\watt}(pulse area $\sim\pi$), the resonance fluorescence signal to background ratio is $\sim 100:1$, Fig.~\ref{fig:scheme}(d). For spectrally-narrower laser pulses, clear Rabi rotations are observed as a function of laser power as demonstrated in Fig.~\ref{fig:scheme}(e).

To calculate the occupation of the trion state, we use the density matrix formalism for a two-level system. We take into account the standard pure dephasing-type coupling to LA phonons via the deformation potential coupling. We note that the phonon coupling in self-assembled QDs is different to that in colloidal QDs where also confined and surface phonons and piezoelectric coupling can play a significant role \cite{Tyagi2010, Sagar2008}. For colloidal QDs, measurements of the phonon dispersion is therefore of crucial importance to understand the coupling. This complication does not arise in our system where dispersion of the LA phonons is well described by a linear relation
 with $\omega_{\mathbf{q}}=c_s |\mathbf{q}|$, with $c_s$ the speed of sound and $\mathbf{q}$ the wavevector. Details on the Hamiltonian and the coupling matrix elements are given in the SI \cite{SI}. The QDs are lens-shaped (Fig.\ \ref{fig:scheme}(b)) with a stronger confinement of the hole than of the electron. We describe this in the calculation with localization lengths for the electron (hole) in the growth direction, $a_{e/h,z}$, and larger localization lengths in the $(x,y)$-plane, $a_{e/h,r}$, taking parameters known from other experiments on QDs of this type \cite{SI}. Specifically, we take GaAs parameters (see SI \cite{SI}) with $a_{e,z}=1.5$~nm, $a_{e,r}= 5.7$~nm and $a_{h,z}/a_{e,z}=a_{h,r}/a_{e,r}=0.77$. From the Hamiltonian \cite{SI} we set up the equations of motion for the phonon-assisted density matrices, truncate the infinite hierarchy of equations using a fourth order correlation expansion, and then perform a numerical integration \cite{Luker2012,Reiter2014,krugel2006bac}. This method has been shown to produce very reliable results \cite{Luker2012,Reiter2014}. 

\begin{figure}[t]{}
	\centering
	\includegraphics[width=1\columnwidth]{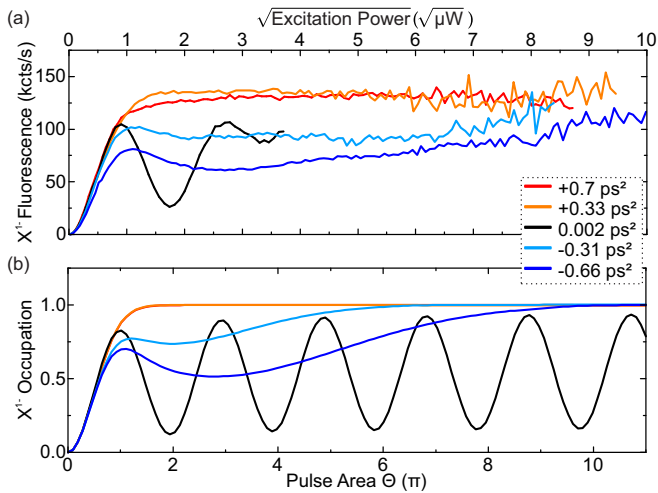}
	\caption{(a) Experimentally measured $\mathrm{X}^{1-}$ resonance fluorescence signal as a function of the square-root of the excitation power, and (b) calculated occupation of the $\ket{\mathrm{X}^{1-}}$ as a function of pulse area for different chirp parameters as indicated.
	}
	\label{fig:fig2}
\end{figure}

\begin{figure}[t]{}
	\centering
	\includegraphics[width=1\columnwidth]{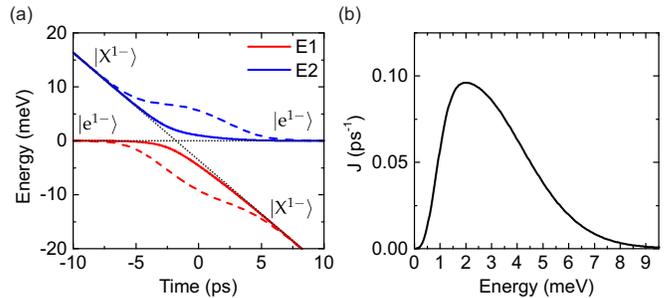}
	\caption{(a) Time evolution of the instantaneous eigenenergies of the coupled electronic-light system for a chirp of $\alpha=0.3$~ps$^2$ and pulse area $\Theta=3\pi$ (solid lines) and $\Theta=10\pi$ (dashed lines). The black dashed lines indicate the uncoupled energies. (b) The phonon spectral density.
	}
	\label{fig:dressed}
\end{figure}

Fig.~\ref{fig:fig2}(a) shows the total $\mathrm{X}^{1-}$ resonance fluorescence signal as a function of the square-root of the excitation power for different chirp values. A Rabi rotation is observed for the smallest chirp: this data set is important to establish the power corresponding to a pulse area of $\pi$. At much higher chirps we enter the RAP regime. We concentrate first on positive chirps. For both \SI{+0.7}{\pico\second\squared} (red curve) and \SI{+0.33}{\pico\second\squared} (orange curve), the signal starts with a fast rise, then saturates and stays nearly constant over the whole excitation power range. It is possible to reach pulse areas up to $11\pi$ on account of the excellent discrimination between reflected laser light and resonance fluorescence signal. The experimental data with positive chirp reflects RAP of a two-level system as described by the detuning dependence of the dressed eigenenergies, Fig.\ \ref{fig:dressed}(a). Starting in the ground state $\ket{\mathrm{e}^{1-}}$, the system evolves along the lower (red) branch and follows the avoided crossing induced by the interaction with the light field provided the pulse area is above the threshold for RAP. After the pulse, the system ends up in the excited state $\ket{\mathrm{X}^{1-}}$. 

The signal measured with negative chirp, Fig.~\ref{fig:fig2}(a), shows a quite different course. Initially, the signal rises, reaching 75\% or 60\% of saturation for a chirp coefficient $\alpha$ of \SI{-0.31}{\pico\second\squared} (cyan curve) or \SI{-0.66}{\pico\second\squared} (blue curve), respectively. Subsequently, the signal decays into a broad minimum, followed by a rise at much higher excitation powers of around \SI{7}{\ensuremath{\sqrt{\micro\watt}}}. At the highest excitation powers, the signal even reaches the signal for a positive chirp. 

We interpret these features as a consequence of the electron-phonon interaction. Phonons can interrupt the adiabatic transfer by causing a jump from one branch to the other, Fig.~\ref{fig:dressed}(a) \cite{Luker2012}. For positive chirp, a phonon can be absorbed taking the system from the lower branch to the upper branch. However, at $T=\SI{4.2}{\kelvin}$, phonons with the required frequency for the transition between the two branches are largely frozen out and the probability for absorption is therefore small. Hence the process with positive chirp is barely influenced by phonons. For negative chirp, in terms of the eigenenergies, the time axis in Fig.\ \ref{fig:dressed}(a) is effectively reversed: the system follows the upper (blue) branch from right to left as the pulse evolves. The system can emit a phonon and jump from the upper to the lower branch. This yields an asymmetry of the RAP with respect to the sign of the chirp \cite{Luker2012,Mathew2014,Glassl2013}. However, the recovery of the RAP signal in the case of negative chirp and large pulse areas suggests that phonon emission, strong for intermediate pulse areas, is suppressed.

The interpretation of the experimental RAP data in terms of phonon scattering is confirmed by theoretical calculations, Fig.\ \ref{fig:fig2}(b). Using pulse parameters from the experiment and a lens-shaped QD geometry leads to good agreement between experiment (Fig.\ \ref{fig:fig2}(a)) and theory (Fig.\ \ref{fig:fig2}(b)). The theoretical results show ideal RAP for positive chirp, rather insensitive to the exact amount of chirp. For negative chirp, a reduced RAP fidelity at intermediate pulse areas and a recovery of the RAP at the highest pulse areas is obtained. Exactly as in the experiment, a stronger negative chirp leads to a stronger reduction of the RAP fidelity at intermediate pulse areas as well as a deferred recovery.
 The detailed agreement with the experimental data shows that phonon scattering is the major factor in the experiment, and in particular, the claim that phonon scattering is suppressed at the highest pulse areas is given strong support by the theory. 
RAP recovery is observed at slightly lower pulse areas in theory as compared to experiment. The reason is not known precisely. A slight increase in chirp at highest powers on account of non-linear effects in the optical fiber may play a role.

In contrast to previous studies \cite{Simon2011,Wu2011,Luker2012,Glassl2013}, the phonons are most efficient at rather low pulse areas for our pulse parameters. This experimental result is in excellent agreement with the theoretical predictions, Fig.~\ref{fig:fig2}(b), which also show a minimum trion population at low pulse areas (around $\sim 2\pi$ for $\alpha=-0.31$~ps$^2$; around $3\pi$ for $\alpha=-0.66$~ps$^2$). This brings the reappearance regime within reach, achieved here above pulse areas of $\sim 8\pi$.   

To interpret our observations in terms of the electron-phonon interaction, we consider the spectral density of the phonons, Fig.~\ref{fig:dressed}(b). The phonon spectral density $J(\omega)$, a measure of the coupling strength between the electron and phonon system at a given frequency $\omega$ \cite{Reiter2014,wigger2014ene,machnikowski2004res}, is defined as $J(\omega)=\sum\limits_{\mathbf{q}} |g_{\mathbf{q}}|^2 \delta(\omega-\omega_{\mathbf{q}})$ with $g_{\mathbf{q}}$ the exciton-phonon coupling matrix element (see SI \cite{SI}). The non-monotonic behavior found in Fig.~\ref{fig:dressed}(b) results from a combination of the momentum dependence of the bulk coupling matrix element, which results in a cubic rise of the spectral density, and the influence of the envelope wave functions of electron and hole, which decouple phonons with wavelengths much smaller than the QD size. The net result is that the phonon spectral density has a broad maximum at phonon energies around $2-3$~meV. Coming back to the dressed states shown in Fig.~\ref{fig:dressed}(a), the phonon emission rate from the upper to the lower branch is proportional to the spectral density at the given energy separation. On examining the dressed state energies we see that for a small pulse area of $\Theta=3\pi$ (solid lines), the splitting between the states is of the order of a few meV and hence the phonons are effective. However, for a larger pulse area of $10\pi$ (dashed lines), the splitting between the states is always above $7$~meV, and in this regime the phonon spectral density is almost zero. Accordingly, the phonons do not affect the RAP for high pulse areas. Indeed, the decoupling from the phonons seen in the theoretical curves above $8\pi$ is also observed in the experimental data, where the signals for positive and negative chirps merge. 

\begin{figure}[t]{}
	\centering
	\includegraphics[width=1\columnwidth]{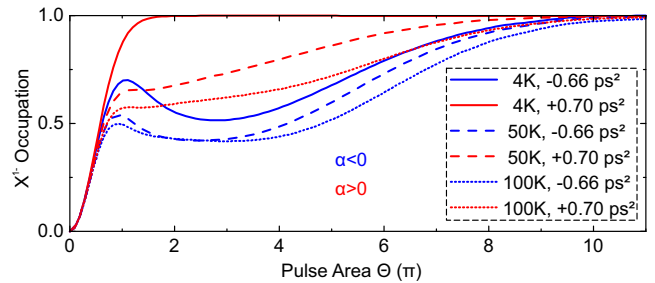}
	\caption{Calculation of the $\ket{\mathrm{X}^{1-}}$ occupation for temperatures of 4~K, 50~K and 100~K (solid, dashed, dotted lines) as a function of pulse area for positive chirp $\alpha=\SI{+0.70 }{\pico\second\squared}$ (red lines) and negative chirp $\alpha=\SI{-0.66 }{\pico\second\squared}$ (blue lines).
	}
	\label{fig:temp}
\end{figure}

We compare our findings with the reappearance phenomena for Rabi rotations \cite{vagov2007non}. For Rabi rotations with typical used pulses, the reappearance only occurs at extremely high pulse areas and has therefore not yet been observed experimentally.
The reason is that for unchirped resonant excitation, the dressed states are degenerate before and after the pulse and they split in the presence of the pulse. Therefore, even in the case of very high pulse areas, the dressed state energy splitting matches the energy at which the phonon spectral density has its maximum in the leading and the trailing edges of the pulse. This explains why the reappearance is more pronounced for hypothetical rectangular pulses \cite{Glassl2011} than for the smooth pulses used experimentally. In contrast, in RAP the dressed states are strongly separated before and after the pulse and, for high pulse areas, never enter the region of efficient phonon coupling. This allows us to enter a
reappearance regime for RAP.

Finally, we estimate how our preparation scheme is affected by elevated temperatures. At higher temperatures, phonons can be both absorbed and emitted. For RAP, this weakens the asymmetry of the phonon influence regarding the sign of the chirp. For example, at $T=\SI{100}{\kelvin}$, phonon scattering limits the exciton population to about $0.55$ for parameters used in Ref. \cite{Luker2012}. Also for the Rabi rotations, a stronger dephasing with increasing temperature has been found experimentally \cite{Ramsay2010}. In the phonon-assisted state preparation scheme \cite{Quilter2015pho}, elevated temperatures are also detrimental for high fidelity preparation. We show the effect of temperature on using the spectrally broad pulses in Fig.~\ref{fig:temp}, where we calculated the occupation of the excited state at a temperature of $T=\SI{50}{\kelvin}$ and $T=\SI{100}{\kelvin}$. For the excitation with positive chirp, the influence of phonons is clearly visible. Instead of immediately rising to an occupation of one, the occupation now goes up to about $0.5$ and then increases gradually. Here, phonon absorption hinders the RAP process by inducing transitions from the lower to the upper branch. However, at high pulse areas above about $8\pi$, we find that the occupation rises to one such that even at these elevated temperatures the electron-phonon scattering is inefficient. The influence of temperature on the exciton occupation in the case of negative chirp is similar. While the damping of the exciton occupation for intermediate pulse areas is increased with respect to 4 K, for higher pulse areas also for negative chirps the occupation goes back to one. Phonon emission as well as the phonon absorption is inhibited in this regime. Hence, by entering the reappearance regime, we can achieve a state preparation scheme robust not only against fluctuations in excitation parameters such as chirp coefficient, detuning and pulse area, but also robust against elevated temperatures.

In conclusion, we have studied the influence of phonons on RAP in the optical domain on a single QD. By performing RAP with highly chirped, spectrally broad laser pulses combined with resonance fluorescence detection together with a full microscopic calculation, we showed that we were able to enter the reappearance regime in which exciton state preparation is minimally influenced by phonon scattering. The work predicts that state preparation in the reappearance regime is almost unaffected by elevated temperatures, a notable feature with respect to other preparation protocols. Our work opens up a new regime for coherent control of excitons in semiconductors with minimal influence from the phonons. 

\begin{acknowledgments}
We acknowledge financial support from EU FP7 ITN S$^{3}$NANO, NCCR QSIT and SNF project 200020\_156637. AL and ADW acknowledge gratefully support from DFH/UFA CDFA05-06, DFG TRR160 and BMBF Q.com-H 16KIS0109.
\end{acknowledgments}


%

\end{document}



\title{Demonstrating the decoupling regime of the electron-phonon interaction in a quantum dot using chirped optical excitation: Supplementary Information}



\author{Timo Kaldewey}
\affiliation{Department of Physics, University of Basel, Klingelbergstrasse 82, CH-4056 Basel, Switzerland}

\author{Sebastian L\"{u}ker}
\affiliation{Institut f\"{u}r Festk\"{o}rpertheorie, Universit\"{a}t M\"{u}nster, Wilhelm-Klemm-Strasse 10, D-48149 M\"{u}nster, Germany}

\author{Andreas V.\ Kuhlmann}
\affiliation{Department of Physics, University of Basel, Klingelbergstrasse 82, CH-4056 Basel, Switzerland}
\affiliation{IBM Research-Zurich, S\"{a}umerstrasse 4, CH-8803 R\"{u}schlikon, Switzerland}

\author{Sascha R.\ Valentin}
\affiliation{Lehrstuhl f\"{u}r Angewandte Festk\"{o}rperphysik, Ruhr-Universit\"{a}t Bochum, D-44780 Bochum, Germany}

\author{Jean-Michel Chauveau}
\affiliation{ Universit\'{e} C\^{o}te d'Azur, CNRS, CRHEA, France}

\author{Arne Ludwig}
\affiliation{Lehrstuhl f\"{u}r Angewandte Festk\"{o}rperphysik, Ruhr-Universit\"{a}t Bochum, D-44780 Bochum, Germany}

\author{Andreas D.\ Wieck}
\affiliation{Lehrstuhl f\"{u}r Angewandte Festk\"{o}rperphysik, Ruhr-Universit\"{a}t Bochum, D-44780 Bochum, Germany}

\author{Doris E. Reiter}
\affiliation{Institut f\"{u}r Festk\"{o}rpertheorie, Universit\"{a}t M\"{u}nster, Wilhelm-Klemm-Strasse 10, D-48149 M\"{u}nster, Germany}

\author{Tilmann Kuhn}
\affiliation{Institut f\"{u}r Festk\"{o}rpertheorie, Universit\"{a}t M\"{u}nster, Wilhelm-Klemm-Strasse 10, D-48149 M\"{u}nster, Germany}

\author{Richard J.\ Warburton}
\affiliation{Department of Physics, University of Basel, Klingelbergstrasse 82, CH-4056 Basel, Switzerland}



\date{\today}



\maketitle 

\beginsupplement

\section{Methods}

\subsection{The quantum dot sample}
\begin{figure}[b]{}
	\centering
	\includegraphics[scale=1]{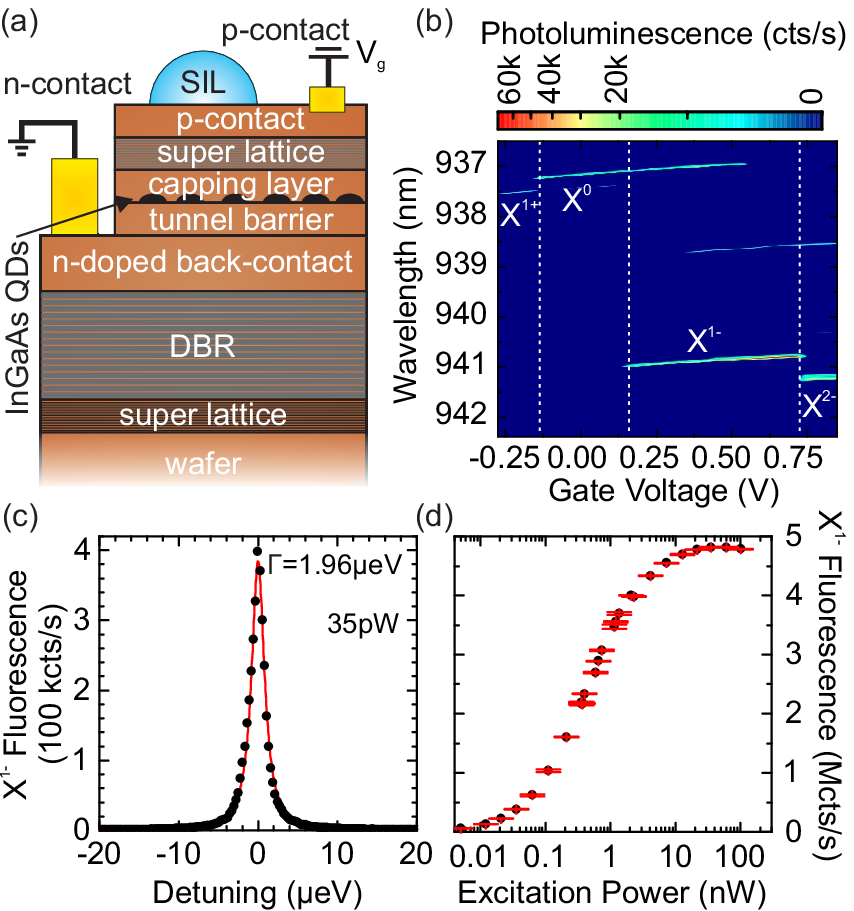}
	\caption{
	\textbf{Resonance fluorescence spectroscopy on a quantum dot (QD) with narrowband continuous wave excitation.}
	\textbf{(a)} the n-i-p layer structure.
	\textbf{(b)} Non-resonant photoluminescence of the QD as a function of the gate voltage $V_g$.
	\textbf{(c)} Narrowband, resonant excitation performance of the QD. An example spectrum showing the resonance fluorescence signal as function of laser detuning at an excitation power of \SI{35}{\pico\watt}. A Lorentzian fit (red solid line) to the black data points shows a FWHM of \SI{1.96}{\micro\electronvolt}.
	\textbf{(d)} Resonance fluorescence signal with resonant continuous wave excitation as a function of the excitation power.
	}
	\label{fig:sfig1}
\end{figure}

We study in this work single self-assembled InGaAs quantum dots (QDs) embedded in a GaAs heterostructure, a diode grown by molecular beam epitaxy. The diode is formed by an n-i-p structure \cite{Prechtel2016}, Fig.\ \ref{fig:sfig1}(a), where the top- and back-gates are epitaxial layers of GaAs doped with carbon and silicon, respectively. The n-i-p diode includes also a distributed Bragg reflector (DBR) which increases the photon collection efficiency. The complete layer sequence is given in Tab.\ \ref{tab:nip}.

The main advantage of the n-i-p diode over an n-i-Schottky diode is an enhanced photon collection by roughly one order of magnitude. On the one hand, this is achieved by the DBR which reflects downward emitted photons towards the objective. On the other hand, the absorption of photons in the top-gate is reduced by using an epitaxial gate instead of a metal gate. 

The final layer of \SI{44}{\nano\meter} undoped GaAs places the p-doped layer, the top-gate, around a node-position of the standing electromagnetic wave in the n-i-p diode. 
The n- and p-doped layers are contacted independently. Selective etching of the capping allows the buried p-layer to be contacted. Access to the n-layer is ensured by wet etching a mesa structure. 

To increase the collection efficiency further we placed a zirconia (ZrO$_{2}$) hemispherical solid-immersion lens (SIL) on the semiconductor surface. The SIL has a refractive index of $n=2.13$ at \SI{940}{\nano\meter}.

In the experiment the n-contact is grounded and a gate voltage $V_g$ is applied to the p-contact. The gate voltage allows control over the charge: single electrons can be loaded into the QD; furthermore, excitonic resonances can be shifted by the DC Stark effect \cite{Warburton2000,Dalgarno2008,Warburton2013}. Fig.\ \ref{fig:sfig1}(b) shows the photoluminescence from a single QD following non-resonant excitation at \SI{830}{\nano\meter} (excitation into the wetting layer). Several charging plateaus are clearly visible in the photoluminescence. They correspond to optical transitions from an excited state to a ground state with different charge: $\ket{\mathrm{X}^{1+}}\rightarrow\ket{\mathrm{h}^{1+}}$, $\ket{\mathrm{X}^0}\rightarrow\ket{0}$, $\ket{\mathrm{X}^{1-}}\rightarrow\ket{\mathrm{e}^{1-}}$ and $\ket{\mathrm{X}^{2-}}\rightarrow\ket{\mathrm{e}^{2-}}$, with $\ket{\mathrm{h}^{1+}}$ and $\ket{\mathrm{e}^{1-}}$ representing a single hole and single electron, respectively. The DC Stark shift is from red to blue with increasing $V_g$ and can be seen clearly. 
We note that the X$^0$ plateau overlaps with the X$^{1-}$ plateau only with non-resonant excitation. We attribute this overlap to  the occasional decay of the $\ket{\mathrm{X}^{1-}}$ via an Auger process \cite{Kurzmann2016}.

\subsection{Resonance fluorescence with continuous wave excitation}

In resonance fluorescence spectroscopy we investigate the QD characteristics with a narrow band (linewidth in sub \si{\pico\meter} range) excitation laser.
The scattering induced by resonant excitation is detected. Back-reflected laser light is suppressed with a dark-field technique leading to an extinction of $10^7$:1 \cite{Kuhlmann2013_NatPhys,Kuhlmann2013_RSI}. A typical $\mathrm{X}^{1-}$ spectrum from the QD at low excitation power is depicted in Fig.\ \ref{fig:sfig1}(c). The linewidth, measured here slowly, is around \SI{2}{\micro\electronvolt} well below saturation, approximately 2.5 times larger than the transform limit \cite{Kuhlmann2013_NatPhys}. A signal to background ratio of more than 1,000: is achieved with linearly polarized excitation. With increasing excitation power the linewidth broadens (power broadening) and the resonance fluorescence signal increases reaching a saturation count-rate of \SI{5}{\mega cts\per\sec} (\SI{5}{\mega\hertz}), Fig.\ \ref{fig:sfig1}(d). This signal is about 10 times more than from QDs in typical n-i-Schottky samples \cite{Kuhlmann2013_NatPhys}. The signal is measured with a single photon avalanche photodiode (SPAD) with an efficiency of 20\% at these wavelengths. Throughout, the stated count-rates are the bare count rates and are not corrected for instance for the poor SPAD quantum efficiency. The photon extraction efficiency from single QDs in the n-i-p device is around 10\%, a success resulting from several improvements in the sample design as discussed above: the n-i-p type features a DBR and a transparent epitaxial gate.

\begin{table}[t]
	\begin{tabular}{llr}
	\hline \hline
	Function & Material & Thickness \\ \hline
	wafer & GaAs &\\
	buffer & GaAs & \SI{50}{\nano\meter}\\
	superlattice, 18 periods & GaAs, AlAs & \SI{72}{\nano\meter}\\
	DBR, 16 periods & GaAs, AlAs & \SI{2400}{\nano\meter}\\
	spacer & GaAs & \SI{57.3}{\nano\meter}\\
	electron reservoir, back contact & GaAs:Si & \SI{50}{\nano\meter}\\
	tunnel barrier & GaAs & \SI{30}{\nano\meter}\\
	InGaAs QDs & InAs & $\sim$1.6 ML\\
	capping layer & GaAs & \SI{153}{\nano\meter}\\
	blocking barrier, 46 periods & AlAs, GaAs & \SI{184}{\nano\meter}\\
	p-doped top contact & GaAs:C & \SI{30.5}{\nano\meter}\\
	undoped spacer & GaAs & \SI{1}{\nano\meter}\\
	etch stop & AlAs & \SI{2}{\nano\meter}\\
	top capping layer & GaAs & \SI{44}{\nano\meter}\\
		\hline \hline
	\end{tabular}
	\caption{Layers of the n-i-p diode. QDs are formed by depositing 1.6 monolayers (ML) of InAs. The distributed Bragg reflector (DBR) increases the photon extraction efficiency.}
	\label{tab:nip}
\end{table}

\subsection{Resonance fluorescence with pulsed excitation}

\begin{figure}[b]{}
	\centering
	\includegraphics[scale=1]{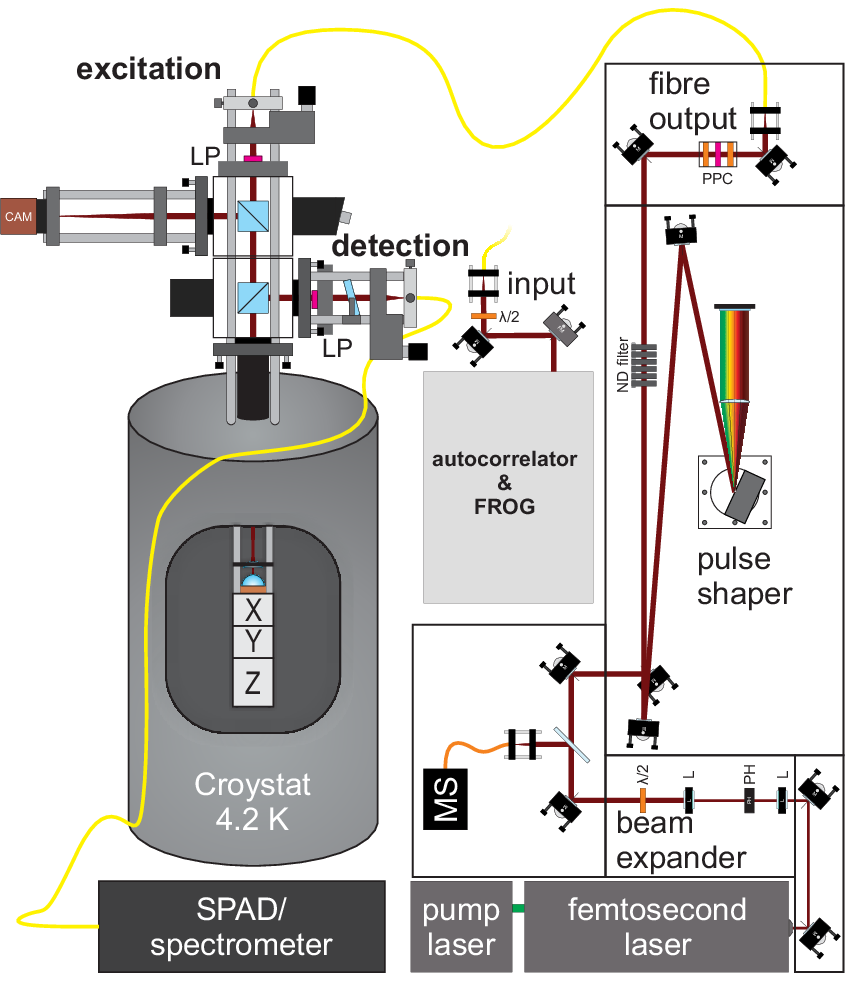}
	\caption{
	\textbf{Scheme of the complete set-up.} Ultra-short pulses were manipulated in a pulse-shaper, coupled into a single mode glass fiber and sent to a confocal microscope mounted on a bath cryostat. The pulses were characterized with a combination of autocorrelator and frequency-resolved-gating (FROG). The sample is held at a temperature of \SI{4.2}{\kelvin} in the bath croystat on an xyz-positioner. A mini spectrometer (MS) measures the spectral profile of the pulses.
	}
	\label{fig:sfig2}
\end{figure}

With pulsed excitation, we use the full spectrum of \SI{130}{\femto\second} transform-limited ``ultra-fast" pulses, again exciting the QD resonantly. The spectral full-width-at-half-maximum (FWHM) of the pulse is $\Delta \lambda=\SI{10}{\nano\meter}$. With this bandwidth we address the ground state transition for the particular charge state set by the gate voltage (but not higher energy transitions). The resonance fluorescence of the QD is detected with a grating spectrometer.

A scheme of the complete set-up is shown in Fig.\ \ref{fig:sfig2}. The ultra-fast pulses from a mode-locked femtosecond laser are expanded and sent into a compact pulse-shaper. The pulse-shaper retains all the spectral components and controls the amount of chirp as described below. From here the pulses pass through power control and polarization optics and are then coupled into a single mode optical fiber. The fiber transports the pulses to a confocal microscope. In the microscope, the excitation pulses pass through a linear polarizer, two beam-splitters and are sent to the objective at cryogenic temperature (4.2 K). The objective, an aspherical lens with a numerical aperture of 0.68, focuses the light onto the sample. The sample is held on a stack of piezo-steppers which allows a particular QD to be placed within the focal spot of the microscope. This process is aided by the in situ diagnostics provided by a camera image of the focus. Light scattered by the QD is collected and coupled into the detection fiber. The back-reflected laser light is suppressed by a second linear polarizer whose axis is orthogonal to the axis of the polarizer in the excitation stage. The resonance fluorescence signal is detected directly with a SPAD (continuous wave excitation) or with a spectrometer-CCD camera (pulsed excitation).

\begin{figure}[b]{}
	\centering
	\includegraphics[scale=1]{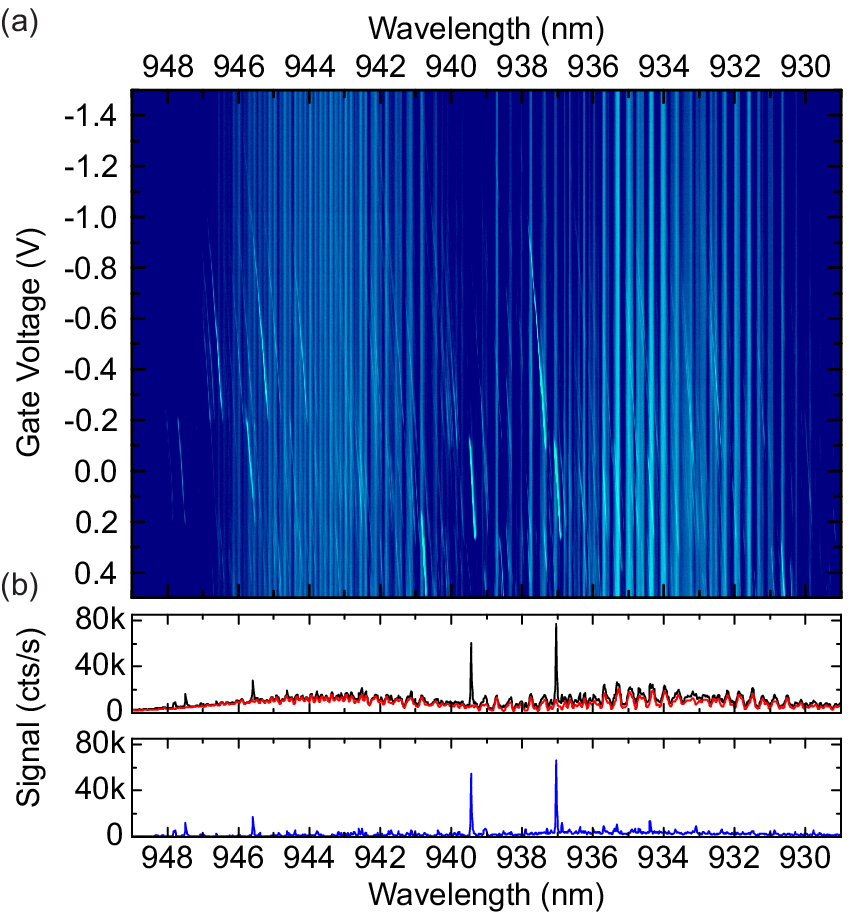}
	\caption{
	\textbf{Laser background subtraction in resonance fluorescence spectroscopy using broadband, pulsed excitation.}
	\textbf{(a)} The resonance fluorescence signal as a function of gate voltage without background subtraction measured with an excitation power of \SI{\sim8}{\micro\watt} corresponding to a pulse area of around 3 $\pi$.
	\textbf{(b)} One spectrum from (a) at a gate voltage of \SI{-0.012}{\volt} is shown as the black curve in the upper panel. Distinct peaks from the exciton and trion are visible toegther with a broad laser background. The reconstructed laser background (from all spectra in (a)) is shown in red. The blue curve in the lower panel shows the difference between the black and the red curve, the resulting signal after background subtraction.
	}
	\label{fig:sfig3}
\end{figure}

Detecting resonance fluorescence with pulsed excitation also depends on suppressing the reflected laser light with the polarization-based dark-field technique. However, owing to the wavelength dependence of the polarization optics, the laser suppression of a broadband pulsed laser is less effective than with the narrowband continuous wave laser: we reach an extinction ratio of typically $10^5$:1 with broadband excitation. We use in addition the spectral mismatch between the broadband laser and the narrowband resonance fluorescence to increase the extinction ratio, Fig.\ \ref{fig:sfig3}. The laser background falling on the CCD camera does not depend on the gate voltage, Fig.\ \ref{fig:sfig3}(a), such that we can use the gate voltage dependence of the QD emission to construct the laser background spectrum. The resulting residual laser background is shown in Fig.\ \ref{fig:sfig3}(b) in red. The spectral shape is explained by the Gaussian spectral shape of the excitation laser along with the quadratic function of laser suppression: the laser suppression is most effective at its alignment wavelength and then decays quadratically in wavelength superimposed by interference fringes. A typical spectrum measured at $V_g=\SI{-12}{\milli\volt}$ is shown in black. Sharp emission peaks from the exciton and biexciton appear on top of the laser background. The signal after background subtraction is shown in the lower plot in blue. Before background subtraction, the maximum signal to background ratio in a narrow spectral window around the emission line is 22:1.

The measurement of emission linewidths in resonance fluorescence with broadband excitation is limited by the resolution of the spectrometer. Hence, we can only state that these linewidths are below \SI{30}{\micro\electronvolt}, as depicted in the spectrum in Fig.\ 1 of the main paper.

\subsection{Pulse shaping: introducing chirp}\label{sec:pulseshaper}
Chirp is introduced into the transform-limited laser pulses by a compact, folded $4f$ pulse-shaper \cite{Martinez1987,Weiner1992}. The scheme is depicted in Fig.\ 1 of the main paper. The unchirped pulse is diffracted by a high resolution, blazed grating (1,800 grooves per mm) and then focused by a lens onto a mirror positioned in the focal plane. From here the light travels back under a small angle with respect to the diffraction plane allowing a spatial separation of incoming and outgoing pulses. The distance between grating and lens controls the chirp \cite{Martinez1987}. The lens and the mirror are mounted on a translation platform. Moving the platform with respect to the grating changes the chirp: if the distance matches the focal length of the lens the chirp is zero, a larger (smaller) distance leads to negative (positive) chirp. The temporal duration of the pulses exiting the laser is \SI{130}{\femto\second} (intensity FWHM); the temporal duration can be stretched up to \SI{15}{\pico\second} corresponding to a chirp in the range from \SI{-0.7}{\pico\second\squared} to \SI{+0.7}{\pico\second\squared}. 

We characterized the chirp with a combination of a FROG and autocorrelator. This allows us to reveal the presence of any high-order phase terms and prevent them by an optimal adjustment of the pulse-shaper. The chirp is measured both in the free space mode and also after the optical single mode fiber in order to determine the sign of the chirp and to compensate for the chirp introduced by the fiber itself.

\section{Theoretical Model}
For the theoretical calculations we use the standard model of a two-level system coupled to longitudinal phonons. For clarity, we outline here the details of the model.

\subsection{Hamiltonian}
We can divide the Hamiltonian $H$ of the system into four parts with
\begin{align*}
H = H_{\textrm{c}} + H_{\textrm{c-l}} + H_{\textrm{ph}} + H_{\textrm{c-ph}},
\end{align*}
where $H_{\textrm{c}}$ denotes the electronic structure, $H_{\textrm{c-l}}$ describes the carrier-light coupling and $H_{\textrm{ph}}+H_{\textrm{c-ph}} $ the phonon part. 

For the electronic structure we take a two-level system, which consists of the ground state $\ket{\mathrm{e}^{1-}}$ and the single exciton state $\ket{\mathrm{X}^{1-}}$ with the Hamiltonian
\begin{align*}
H_{\textrm{c}} = \hbar\omega_{\mathrm{X}^{1-}}\ket{\mathrm{X}^{1-}}\bra{\mathrm{X}^{1-}},
\end{align*}
where $\hbar\omega_{\mathrm{X}^{1-}}$ denotes the energy of the negative trion. The energy of the ground state has been set to zero.

The carrier-light interaction is modeled in the usual dipole and rotating wave approximation. For this two-level system we have
\begin{equation*}
H_{\textrm{c-l}} = \, \frac{\hbar}{2}\left(\Omega(t)|\mathrm{X}^{1-}\rangle\langle \mathrm{e}^{1-}| + \Omega^*(t)|\mathrm{e}^{1-}\rangle\langle \mathrm{X}^{1-}|\right). \\
\end{equation*}
where $\hbar\Omega(t)=2\mathbf{M}\cdot \mathbf{E}(t)$, with $\mathbf{E}(t)$ the positive frequency component of the electric field of the laser pulse; $\mathbf{M}$ denotes the dipole matrix element. 

The main source of decoherence in QDs is caused by the coupling to longitudinal acoustic (LA) phonons which are treated as bulk-like due to the small acoustic mismatch between the QD area and the surrounding material \cite{Reiter2014}. The Hamiltonian of the free phonons reads
\begin{align*}
H_{\textrm{ph}} = \hbar\sum_{\mathbf{q}}\omega_{\mathbf{q}}b^\dag_{\mathbf{q}}b_{\mathbf{q}},
\end{align*}
where $b^\dag_{\mathbf{q}}$ ($b_{\mathbf{q}}$) is the creation (annihilation) operator of a phonon with wave vector $\mathbf{q}$ and frequency $\omega_{\mathbf{q}} = c_{s}|\mathbf{q}|$, $c_{s}$ being the sound velocity.

The carrier-phonon interaction is modeled by a pure-dephasing Hamiltonian using the deformation potential coupling mechanism. The corresponding Hamiltonian is given by
\begin{align*}
H_{\textrm{c-ph}} = \hbar\sum_{\mathbf{q}} \left(g_{\mathbf{q}}b_{\mathbf{q}} + g^*_{\mathbf{q}}b^\dag_{\mathbf{q}}\right) |\mathrm{X}^{1-} \rangle\langle \mathrm{X}^{1-}| .
\end{align*}
The main coupling mechanism is the deformation potential coupling with the coupling matrix element for the electrons and holes
\[
	g_{\mathbf{q}}^{e/h}=\sqrt{\frac{q}{2V\rho\hbar c_{s}}}D_{e/h}F_{\mathbf{q}}^{e/h},
\]
where $V$ is the normalization volume, $\rho$ the mass density of the crystal, and $D_{e/h}$ the deformation potential coupling constant for electrons/holes. As parameters we take the standard GaAs parameter listed in Tab.~\ref{tab:para}. The exciton coupling matrix element is obtained by $g_{\mathbf{q}} = g_{\mathbf{q}}^{e}- g_{\mathbf{q}}^{h}$.

The form factor $F_{\mathbf{q}}^{e/h}$ accounts for the bound states of the QD and, thus, for the geometry of the QD. We assume a harmonic confinement potential in a lens-shaped QD yielding the form factor
\[
	F_{\mathbf{q}}^{e/h}=\exp\left[-\frac{1}{4}\left(q_z^2 a_{e/h,z}^2+q_{r}^2a_{e/h,r}^2\right)\right].
\]
$q_z$ is the wave vector in the $z$-direction; $q_{r}$ is the in-plane wave vector. $a_{e/h,z}$ and $a_{e/h,r}$ are the electron/hole localization lengths in the $z$-direction and in the $(x,y)$-plane direction, respectively. We used $a_{e,z}=1.5$~nm and $a_{e,z}= 5.7$~nm, thereby modelling a flat QD. The ratio between electron and hole localization lengths is taken to be $a_h/a_e=0.77$ for both directions \cite{warburton2002gia}.

\begin{table}[t]
\begin{tabular}{llr}
\hline \hline
Parameter & & Value \\ \hline
density & $\rho$& $5370\,$kg/m$^3$\\
sound velocity & $c_{s}$ & $5.1\,$nm/ps\\
electron deformation potential constant &$D_e$&$7$~eV\\
hole deformation potential constant &$D_h$&	$-3.5$~eV\\
	\hline \hline
\end{tabular}
\caption{\label{tab:para} Material parameters used in the calculation.}
\end{table}

Before the laser pulse, we assume that the electronic system is in the
ground state and the phonon system is in a thermal equilibrium given by
a Bose distribution at temperature $T$ with $\langle
b^\dag_{\mathbf{q}}b_{\mathbf{q}} \rangle = n_q(T) = 1/(e^{\hbar c_s
q/k_B T}+1)$  with $k_B$ the Boltzmann constant. If not denoted
otherwise, we use liquid helium temperature at $T=4~K$. Assuming an
initially uncorrelated state, we can take the product state between
carrier and phonon system as initial condition for the dynamics.

Using this Hamiltonian we set up the equations of motion in the density matrix formalism which leads to an infinite hierarchy of phonon-assisted variables. We truncate this hierarchy using a fourth-order correlation expansion, which yields reliable results for the carrier-dynamics in a QD \cite{Reiter2014,Luker2012,glassl2011lon}. The equations of motion of the two-level model can be found, e.g., in Ref.~\cite{krugel2006bac}.


%